\documentclass[aps,prl,showpacs,twocolumn,amsmath,amssymb]{revtex4}

\usepackage{amsmath}    
\usepackage{graphicx}   
\usepackage{graphics}
\usepackage{dcolumn}
\usepackage{bm}

\usepackage{multirow}

\usepackage{latexsym}
\usepackage{amssymb}
\usepackage{amsmath}

\usepackage{color}
\usepackage{epstopdf}

\usepackage[normalem]{ulem}

\begin{document}
\title{The hidden order in URu$_2$Si$_2$: Symmetry-induced anti-toroidal vortices}
\author{Vladimir E. Dmitrienko\footnote{email: dmitrien@crys.ras.ru} and Viacheslav A. Chizhikov\footnote{email: chizhikov@crys.ras.ru}}
\affiliation{A.V. Shubnikov Institute of Crystallography, FSRC ``Crystallography and Photonics'' RAS, Leninskiy Prospekt 59, 119333, Moscow, Russia}

\begin{abstract}
We discuss possible approaches to the problem of the URu$_2$Si$_2$
``Hidden Order'' (HO) which remains unsolved after tremendous efforts of
researches. Suppose there is no spatial symmetry breaking at the HO
transition temperature and solely the time-reversal symmetry breaking
emerges owing to some sort of magnetic order. As a result of its $4/mmm$
symmetry, each uranium atom is a {\em three-dimensional} magnetic vortex;
its intra-atomic magnetization $\mathbf{M(r)}$ is intrinsically
non-collinear, so that its dipole, quadrupole and toroidal moments vanish,
thus making the vortex ``hidden''. The first non-zero magnetic multipole
of the uranium vortex is the toroidal quadrupole. In the unit cell, two
uranium vortices can have either the same or opposite signs of
$\mathbf{M(r)}$; this corresponds to either {\em ferro-vortex} or {\em
antiferro-vortex} structures with $I4/mmm$ or $P_{I}4/mmm$ magnetic space
groups, respectively. Our first-principles calculations suggest that the
vortex magnetic order of URu$_2$Si$_2$ is rather strong: the total
absolute magnetization $|\mathbf{M(r)}|$ is about 0.9 $\mu_B$ per U atom,
detectable by neutron scattering in spite of the unusual formfactor. The
ferro-vortex and antiferro-vortex phases have almost the same energy and
they are energetically favorable compared to the non-magnetic phase.
\end{abstract}
\maketitle

\section{Introduction}\label{sec:Introduction}

For more than thirty years, since the first papers appeared in 1985
\cite{Palstra1985,Maple1986,Schlabitz1986}, there were many attempts to
understand the mysterious Hidden Order (HO) in the heavy-fermion compound
URu$_2$Si$_2$ (they are surveyed in two detailed reviews
\cite{Mydosh2011,Mydosh2014}). The main problem is that below the HO
transition temperature, $T_{HO}=17.5$ K, there are practically no obvious
physical phenomena associated with the order parameter; the only
unequivocal evidence for the order is a rather strong specific heat jump
\cite{Palstra1985,Maple1986,Schlabitz1986} at $T_{HO}$. For instance, the
accompanying antiferromagnetic order violating the body-centered symmetry
\cite{Broholm1987,Isaacs1990,Broholm1991,Walker2011} is so weak that it
cannot explain the behavior of the specific heat. The observed lattice
symmetry breaking \cite{Tonegawa2014} and in-plane anisotropy of the
magnetic susceptibility \cite{Okazaki2011} are also extremely weak and
their relation to HO is not clear \cite{Trinh2016}. The symmetry
breaking from body-centered tetragonal to simple tetragonal was carefully
examined \cite{Butch2015} via inelastic neutron and x-ray scattering
measurements, and no signs of reduced spatial symmetry, even in the HO
phase, had been found. The fourfold local symmetry of the HO state of
URu$_2$Si$_2$ has recently been confirmed by means of singlecrystal NMR
measurements \cite{Kambe2018}.

There have been several interesting attempts to understand the HO transition within the phenomenological Landau-Ginzburg theory  (see for instance \cite{Shah2000,Mineev2005,Haule2010,Chandra2015} and references therein). The phenomenological approaches include naturally both the hidden order and the pressure-dependent antiferromagnetic order. They also take into account the results of {\em ab initio} studies. However, for the time being, the problem is still very open, and we also discuss other possible forms of the order parameter in this paper.

A popular idea is that there is time-reversal symmetry breaking (TRSB)
related probably with an exotic type of multipole magnetic order emerging
at $T_{HO}$ \cite{Mydosh2014,Walker1993,Takagi2012}. Many efforts,
theoretical and experimental, were concentrated on searching for possible
multipole orders
\cite{Mydosh2014,Ikeda2012,Khalyavin2014,Suzuki2014,Knafo2017,Wang2017,Suzuki2018}.
However, the conventional methods like magnetic neutron and resonant x-ray
scattering seem unable to detect those multipoles.

In this paper, we suggest a simple HO model based mainly on the symmetry
consideration. Indeed, if we cannot detect any pronounced violation of the
{\em spatial} symmetry below $T_{HO}$, let us assume that the HO has
exactly the same symmetry, namely $4/mmm$, as the high temperature
paramagnetic phase of URu$_2$Si$_2$. More precisely, we suppose that HO is
a non-collinear intra-atomic magnetization of uranium atoms with $4/mmm$
symmetry so that the only symmetry violation at the transition point is
TRSB. Surprisingly, such a simple assumption leads to a non-trivial vortex
HO described by the toroidal quadrupole order parameter which is difficult
to detect with conventional methods. First-principles calculations show
that the vortex HO is perhaps strong enough to be detected by careful
monitoring of neutron reflections across the phase transition.

\section{Magnetic symmetry of the hidden order}\label{sec:symmetry}

We first remind that the magnetic moment $\mathbf{M(r)}$ is a
pseudo-vector and transformations of its components under mirror
reflections are just opposite to a usual vector: the component normal to
the mirror plane keeps its direction whereas the parallel components
invert their directions. For instance, for the $m_z$ mirror plane,
$M_z(x,y,-z)=M_z(x,y,z)$, $M_x(x,y,-z)=-M_x(x,y,z)$, and
$M_y(x,y,-z)=-M_y(x,y,z)$. The space inversion does not change
$\mathbf{M(r)}$: $\mathbf{M(-r)}=\mathbf{M(r)}$. The time reversal
symmetry operation, denoted by the prime sign, inverts the direction of
the magnetization: $\mathbf{M^\prime(r)}=-\mathbf{M(r)}$.

\begin{figure}[h]
\includegraphics[width=7cm]{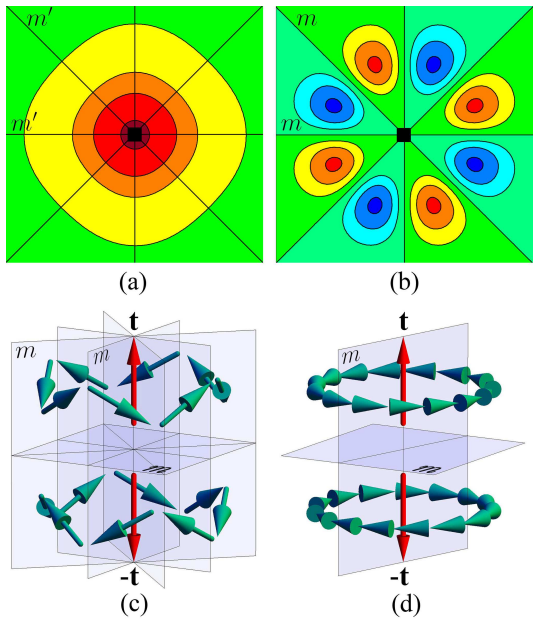}
\caption{\label{figHO}
(Color online) Different symmetries of intra-atomic magnetization of
uranium atoms in URu$_2$Si$_2$. (a) A conventional magnetic atom would
have the point symmetry $4/mm^\prime m^\prime$ including the vertical
fourfold axis (black square), the horizontal mirror plane $m$ (the figure
plane) and two types of vertical pseudo-mirror planes $m^\prime$
(perpendicular to $x$ or $y$ axes and diagonal to them, black lines);
$m^\prime$ means a combine operation of mirror reflection and time
reversal. (b) In-plane uranium magnetization with the $4/mmm$ symmetry
where red and blue colors correspond to positive and negative regions of
$M_z(x,y,z=0)$ divided by mirror planes $m$ (black lines). (c) 3D
Anti-Toroidal Vortex (ATV): sixteen $4/mmm$-equivalent magnetic vectors
(green arrows) form two 8-vector vortices at $\pm z$ with opposite
directions of toroidal moments (red arrows); see also a movie in Ref.~\cite{supplemental1}.
It should be emphasized that the $4/mmm$ symmetry
induces the ATV structure only for pseudo-vectors like $\mathbf{M(r)}$ and
not for true vectors like electric dipole moments, {\em etc}. (d) In
principle, higher symmetries are also possible, up to $\infty/mm$, which
is the symmetry of the nematic order.}
\end{figure}

The principal difference between conventional magnetic atoms and a
magnetic atom with $4/mmm$ symmetry is obvious from Fig.~\ref{figHO}. The
magnetic point symmetry of conventional atoms would be $4/mm^\prime
m^\prime$ and it includes one vertical 4-fold axis, one horizontal mirror
plane $m$ and two types of vertical mirror planes $m^\prime$ (normal and
diagonal to $x,y$ axes). As a result of this symmetry, $M_x(x,y,0)=0$ and
$M_y(x,y,0)=0$ in the $z=0$ mirror plane and usually these components
remain to be small above and below the mirror plane so that the main
magnetization of the atom is $M_z$.

For the case of $4/mmm$ symmetry, $M_x$ and $M_y$ are also zero in the
plane of the figure but $M_z$ should be very inhomogeneous, it should
change its sign at least eight times when we go around the atom
(Fig.~\ref{figHO}b). In the horizontal mirror plane $z=0$, we have eight
similar sectors with alternating $M_z$-component. Then, passing through
all vertical mirror planes, shown in Fig.~\ref{figHO}b by black lines, the
parallel components of $\mathbf{M(r)}$ become zero and change their signs.
In other words, for all \textbf{r} belonging to the mirror planes of
$4/mmm$ symmetry, $\mathbf{M(r)}$ is normal to the corresponding plane.

For any general position $x,y,z$ the $4/mmm$ symmetry operations create a
pair of eight-vector vortices with head-to-tail arrangement of {\em
equivalent} moments in the $\pm z$ planes (Fig.~\ref{figHO}c). The
toroidal moments \cite{Dubovik1990,Spaldin2008,Kopaev2009} of these two
eight-vector vortices are anti-parallel (along $\pm \mathbf{z}$). It is a
general magnetic arrangement dictated by the $4/mmm$ symmetry and it makes
the $\mathbf{M(r)}$ field significantly non-collinear and inhomogeneous
simply as a result of the symmetry. Each uranium atom looks like an
atomic-size magnetic skyrmion built from two equivalent halves at $z>0$ and $z<0$
with opposite toroidal moments. We could refer to this configuration as an
Anti-Toroidal Vortex (ATV). It should be emphasized that the $4/mmm$
symmetry induces the ATV HO only for pseudo-vectors like $\mathbf{M}$ and
not for true vectors like electric dipole moments, etc.

To characterize quantitatively the inhomogeneous atomic magnetization with
$4/mmm$ point symmetry we can use the tensor moments of $\mathbf{M(r)}$
relative to the atomic center. The average dipole moment $\langle
\mathbf{M(r)}\rangle$ is zero. Here and below $\langle \ldots \rangle$
means integration $V^{-1}\int \ldots d\mathbf{r}$ over a spherical
atomic-size volume $V$ around the atom. The magnetic quadrupole moment
$\langle M_{i}(\mathbf{r})x_{j}\rangle$=0 because of the inversion center
$\mathbf{M(-r)}=\mathbf{M(r)}$. In particular, the atomic toroidal
(anapole) moment $\langle [\mathbf{r\times M(r)}]\rangle $
\cite{Dubovik1990,Spaldin2008,Kopaev2009}, which is an antisymmetric part
of this tensor, is zero as well as the monopole moment $\langle
\mathbf{r\cdot M(r)}\rangle $. For the same reason, all even-rank tensor
moments of $\langle M_{i}(\mathbf{r})x_{j}\ldots x_{n}\rangle$ type are
zero as well.

Thus the first non-zero tensor moment of the $4/mmm$ ATV structure is the
third-rank tensor $M_{ijk}=\langle M_{i}(\mathbf{r})x_{j}x_{k}\rangle$; it
is symmetric under permutation of the last two indices. It is easy to show
(or to find in textbooks \cite{Sirotin82}) that for this symmetry the
third-rank pseudo-tensor $M_{ijk}$ has only four non-zero components and
all of them are equal up to the sign:
$M_{123}=M_{132}=-M_{231}=-M_{213}=M_v$, where $M_v=\frac12 V^{-1}\int
[\mathbf{M(r)\times r]\cdot z}d\mathbf{r}$. The time-odd parity-even moment
$M_v$ characterizes the strength and sign of the ATV HO (it is
called either magnetic octopole or quadrupole toroidal moment).

The sign of $M_v$ is a non-trivial attribute. Indeed, we can change the
sign of $M_v$ by reversing the magnetization direction in all points
$\mathbf{M(r)\rightarrow -M(r)}$ (the time reversal operation). However
this way we obtain a new object which cannot be superposed with the old
one neither by rotations nor by mirror reflections. Since the only symmetry
operation relating these two objects is the time inversion, their energies
must be equal. Thus any magnetization arrangement with $4/mmm$ point
symmetry can exist in two energetically equivalent variants with $\pm
M_v$. It is natural to call them clockwise and anticlockwise vortices for
$M_v>0$ and $M_v<0$, correspondingly. However, it should
be emphasized that the sign of $M_v$ is not topologically stable, it can
be changed by deformation of the $\mathbf{M(r)}$ field. The magnetization
arrangement with $M_v=0$ can correspond to non-zero absolute magnetization
$\langle |\mathbf{M(r)}|\rangle\neq 0$.

There are two uranium atoms in the body-centered tetragonal unit cell of
URu$_2$Si$_2$. In the simplest magnetic structure, both atoms have the
vortices with the same $M_v$, either both clockwise or both anticlockwise.
Such structure has $I4/mmm$  magnetic space group \cite{Bilbao,Perez-Mato2015} and can be
called the ferro-vortex phase. The clockwise and anticlockwise
ferro-vortex phases should have equal energies and can be mutually
transformed  by the time reversal. The clockwise and anticlockwise domains
can coexist, being separated by domain walls, in real samples of the
ferro-vortex phase.

If those two atoms have opposite magnetization directions (one clockwise
and another anticlockwise) then the lattice is primitive and the magnetic
symmetry group is $P_{I}4/mmm$ \cite{Bilbao,Perez-Mato2015}. In this case, the lattice
consists of clockwise and anticlockwise layers alternating along the
$z$-axis; it can be called the antiferro-vortex phase. The time reversal
is equivalent to the $(\frac12,\frac12,\frac12)$ shift of the lattice.

Besides ferro-vortex and antiferro-vortex phases many (infinite!)
symmetrically different arrangements of the clockwise and anticlockwise
vortices are possible but their consideration should be left for the
future work. Then, in principle, ATV with higher symmetries are also
possible, up to $\infty/mm$, which is the symmetry of the nematic order
(see Fig.~\ref{figHO}d). An open question is whether the vortices with
such a high symmetry can exist in free atoms, molecules or nematic-like
liquid crystals. Actually, the toroidal quadrupole moments are discussed
for positronium atoms \cite{Porsev1994} and for deuterons
\cite{Mereghetti2013} (a survey of related works is given in
\cite{Gray2010}).

The quantitative characterization of $4/mmm$ vortices by the third-rank
tensor $M_{ijk}$ has three important complications: (i) $M_{ijk}$ does not
depend on the azimuthal orientation of the vortex in the $xy$ plane: (ii)
it does not distinguish between $4/mmm$ and other uniaxial symmetries
($422$, $4mm$, $\bar{4}2m$, $622$, $6mm$, $\bar{6}2m$, $6/mmm$, $\infty2$,
$\infty m$, $\infty/mm$); (iii) the $M_z$ component gives no contribution
to $M_v$. Some of these drawbacks disappear for the next non-zero tensor
(fifth-rank) and for the magnetoelectric tensor $\langle
M_{i}(\mathbf{r})E_{j}(\mathbf{r})x_{k}\rangle$. All this means that pure
symmetrical consideration leaves a lot of freedom for possible scenarios
of the HO transition and more work is needed here.

\begin{figure*}[!ht]
\includegraphics[width=5cm]{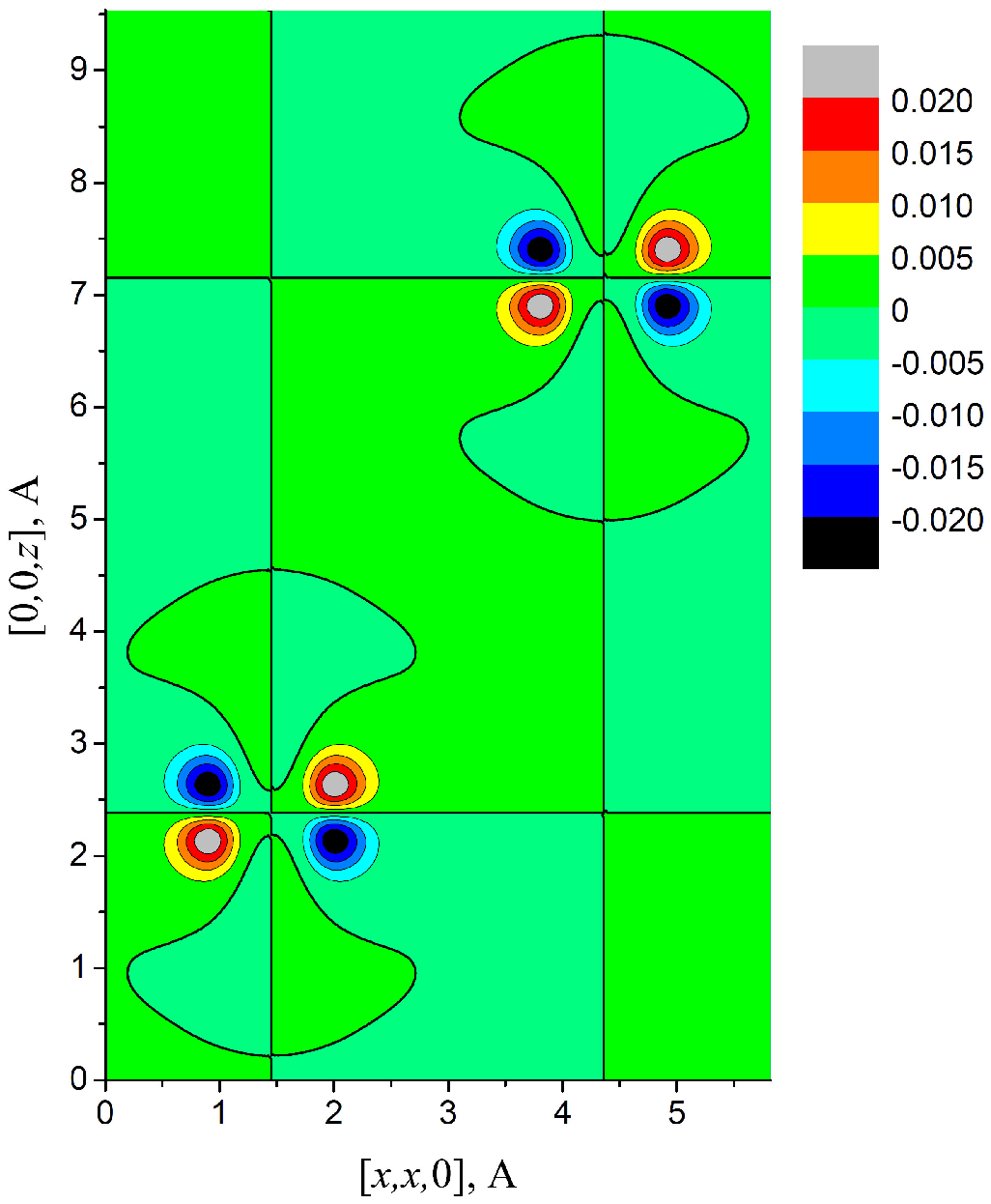}
\includegraphics[width=5cm]{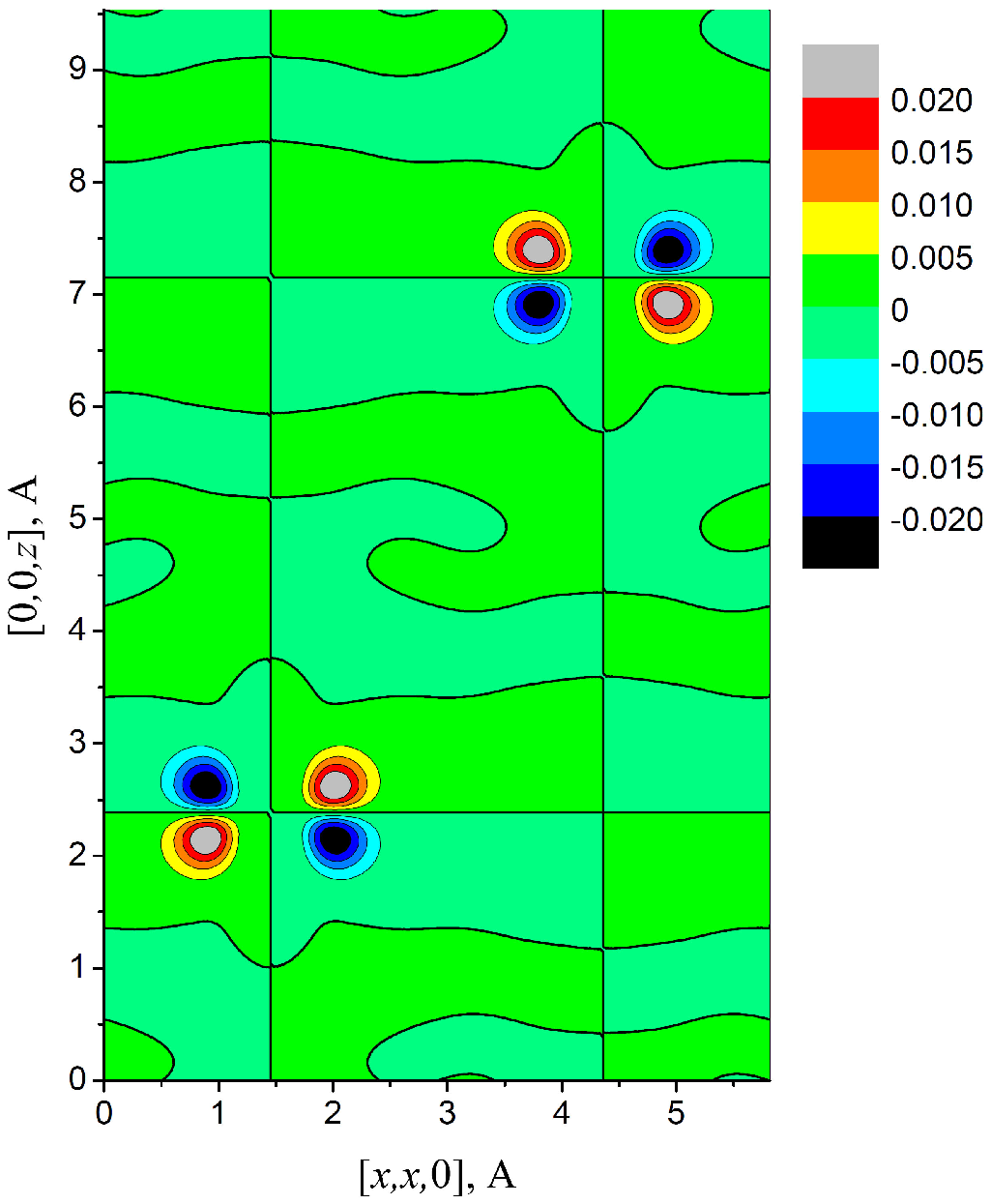}
\includegraphics[width=5cm]{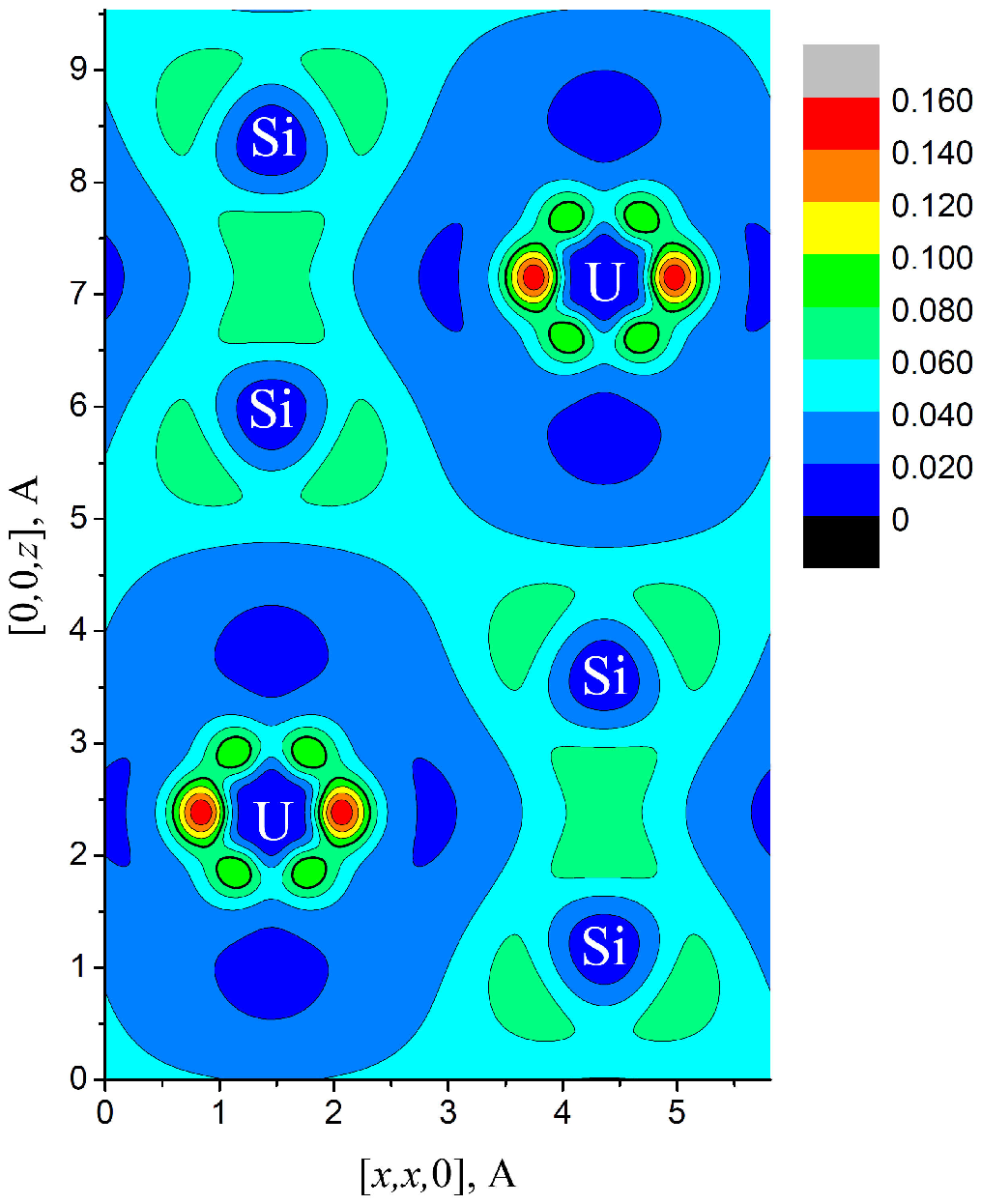}
\caption{\label{diagonalMx} (Color online) The calculated magnetization
distribution $M_x(\mathbf r)$ within the diagonal mirror plane formed by
vectors $[1,1,0]$ and $[0,0,1]$ in the unit cell of the ferrovortex (a)
and antiferrovortex (b) phases; (c) the calculated valence electron
density which is almost equal for both phases. In this plane, $M_y(\mathbf
r)=-M_x(\mathbf r)$ and $M_z(\mathbf r)=0$. Two uranium atoms are at
$(\frac14,\frac14,\frac14)$ and $(\frac34,\frac34,\frac34)$ positions with
Si atoms surrounding them; Ru atoms are out of the plane. The straight
lines are intersections with vertical and horizontal mirror planes where
$M_x$ and $M_y$ change their signs. }
\end{figure*}

\section{First principle simulations}\label{sec:abinitio}

The symmetry-based approach is of course reliable, but it cannot say
whether and when those exotic anti-toroidal vortices could be
energetically stable, what are the values of $\mathbf{M(r)}$ in different
points of the unit cell, {\em etc.} To find the magnetization
$\mathbf{M(r)}$, the electronic densities $\rho(\mathbf{r})$ and the
energies of possible URu$_2$Si$_2$ phases we have performed
``illustrative'' {\em ab initio} simulations using the {\sc Quantum
ESPRESSO} package \cite{QE,QE-2009} with appropriate pseudopotentials and techniques 
\cite{Perdew1992,pw91,Perdew1996,Wu2006,Becke1988,Perdew1986,Perdew1981,Monkhorst,Marzari}.

We do not fix the spatial and magnetic symmetries of URu$_2$Si$_2$ in the
beginning and during the self-consistent minimization procedure. Instead,
the procedure starts from crystal structures whose symmetries are
subgroups of $I4/mmm$. Small initial magnetic moments are assigned to
silicon and ruthenium atoms so that uranium magnetic moments are not
predetermined. Then during the self-consistent iterations those
conventional magnetic moments become smaller and smaller but at the same
time new magnetization field $\mathbf{M(r)}$ (with zero average
magnetization) is growing mainly around uranium atoms, i.e. the absolute
magnetization $\langle |\mathbf{M(r)}|\rangle$ is progressively growing
until an equilibrium structure is reached. Symmetry analysis of the
appearing magnetization shows that new symmetry elements initially look
like some tendency and then become more and more exact if the iterative
self-consistent procedure converges. See Ref.~\cite{supplemental2} for more
details of the simulations.

Both the ferro-vortex and antiferro-vortex phases have been obtained in
our simulations starting from different initial structures. Their energies
are well below the energy of non-magnetic phase: per formula unit, $\Delta
E_{fv}$=-0.0318 eV/f.u. and $\Delta E_{av}$=-0.0364 eV/f.u. This energy
gain seems to be too strong for the observed value
\cite{Palstra1985,Maple1986,Schlabitz1986} of the specific heat jump
corresponding to the internal energy change induced by the hidden order of
about 0.00018 eV/f.u. In fact, the energy responsible for the HO phase
transition is of about an {\em interaction energy} between magnetic atoms,
which is ``fighting'' with entropy for the phase transition. The
interaction energy is a very small part of the total magnetic energy and
the former is impossible to extract from the latter within the
conventional DFT simulations. Quite probably, the anti-toroidal vortices
appear as fluctuations well above the HO transition temperature, and they
are arranged into ferro-vortex or antiferro-vortex phase at the HO
transition temperature owing to very subtle interactions between vortices.

The calculated magnetization and charge densities are shown in
Fig.~\ref{diagonalMx} for the diagonal mirror plane $x=y$ including two U
atoms. The main magnetic and charge features obviously correspond to the
$5f$ uranium orbitals \cite{Kvashnina2017} (mean radius 0.76 \AA). The
uranium vortices are almost the same for both phases, except that in the
antiferro-vortex phase they have opposite signs. And the total absolute
magnetization is almost the same for both phases:
$|\mathbf{M(r)}_{fv}|$=0.93 $\mu_B$/f.u. and $|\mathbf{M(r)}_{av}|$=0.96
$\mu_B$/f.u. According to Ref. \cite{Amitsuka2002}, the value of about 1
$\mu_B$/f.u. is needed to explain the observed specific heat jump. The
magnetization is concentrated around uranium atoms
(Fig.~\ref{diagonalMx}a,b): in the ferro-vortex(antiferro-vortex) phase,
there is about 0.936 (0.93) of the total $|\mathbf{M(r)}|$ inside the
Slater uranium radii ($R_{\mbox{s}}$=1.75 \AA) and remaining itinerant
magnetization is distributed in the unit cells according to their space
symmetries. The very strong anisotropy of ATVs could naturally explain the
Ising-like behavior of HO \cite{Mydosh2014,Trinh2016}. The calculated
$M_z$ for one atom is shown in Fig. \ref{figHO}b which is a $0.5\times
0.5$ part of the unit cell $xy$ plane (i.e. about $2\times 2$ \AA$^2$);
see also movies in Ref.~\cite{supplemental3}.

In fact, it is well known that magnetic intra-atomic non-collinearity is a
general effect, arising because of the relativistic spin-orbit coupling
not only in actinides \cite{Nordstrom1996} but also in other materials
\cite{Sandratskii1998,Bultmark2009,Ma2015}. The non-collinear magnetism is
very sensitive to the space group symmetry and we have predicted recently
\cite{Tsvyashchenko2016} the toroidal intra-atomic moments for RhGe
crystal with the $P2_13$ space group. The case of URu$_2$Si$_2$ is
especially interesting because its symmetry is so high that observation of
its intra-atomic vortices is really a non-trivial problem.

\begin{figure}[!hb]
\includegraphics[width=3.5cm]{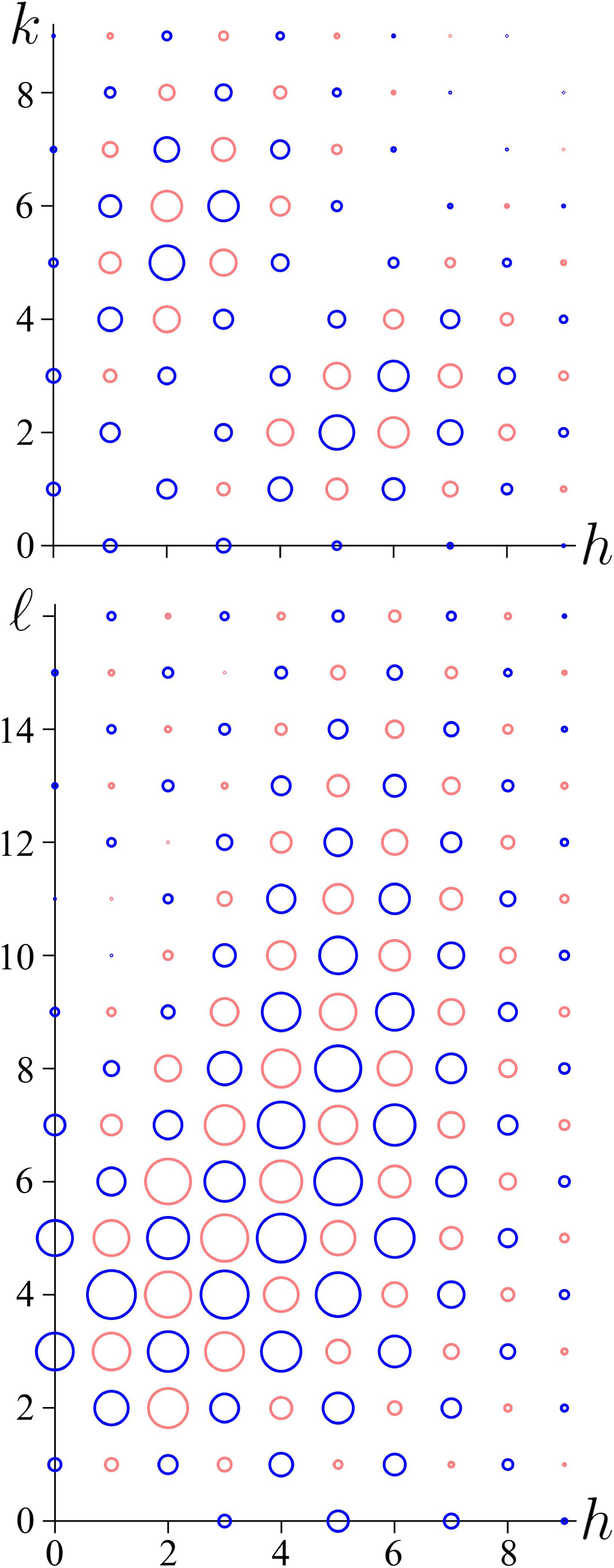}
\includegraphics[width=3.5cm]{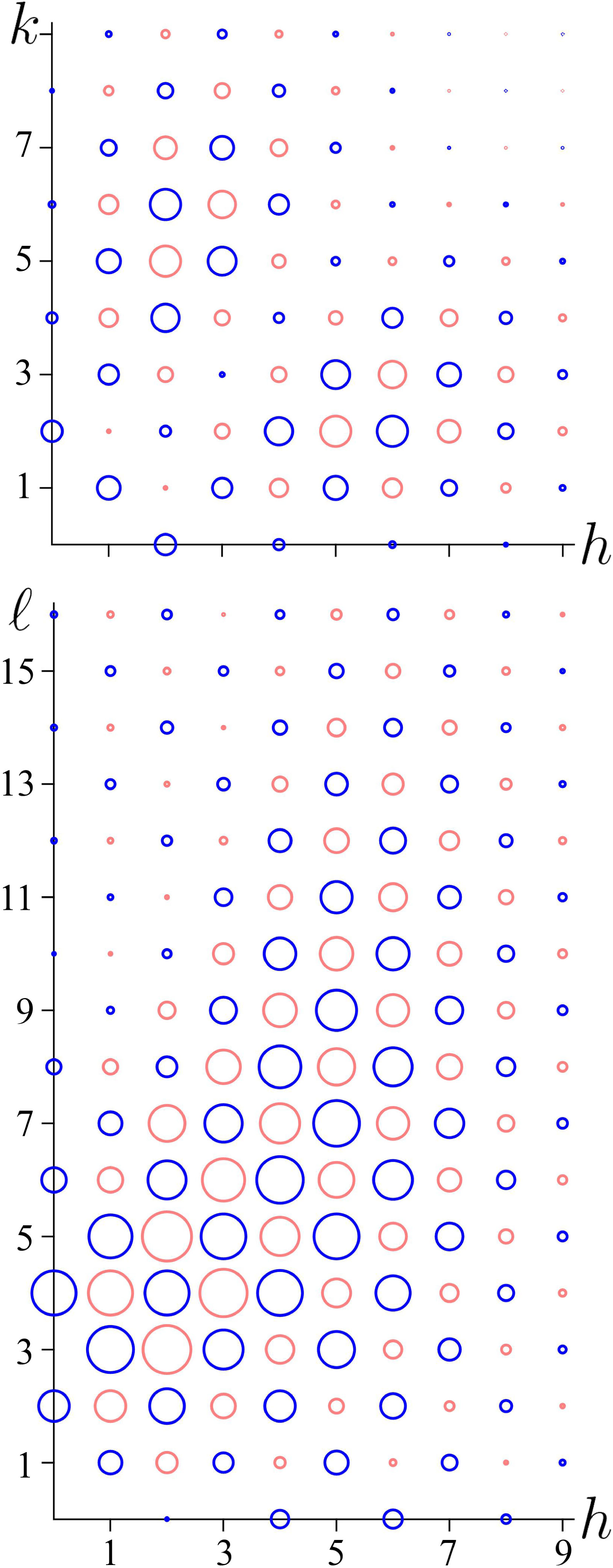}
\caption{\label{Bragg} (Color online) Relative intensities of magnetic
contributions to neutron reflections for the ferro-vortex ($h+k+\ell=2n$,
left) and antiferro-vortex ($h+k+\ell=2n+1$, right) phases; the circle
areas are proportional to $|\mathbf{M}(hk\ell)|^2$ and normalized on the
most intense magnetic reflections, 255 and 256 for the ferro-vortex and
antiferro-vortex phases, respectively. Top: for $hk0$ (red) and $hk1$
(blue). Bottom: for $h0\ell$ (red) and $h1\ell$ (blue). The anti-toroidal
magnetization of uranium atoms (Fig. \ref{diagonalMx}a,b) results in a
rather unusual reciprocal-space distribution of strong reflections:
magnetic contributions are zero for $h00$, $0k0$, $00\ell$, and $hh0$
reflections.}
\end{figure}

\section{Discussion}\label{sec:discussion}

The logic of our approach is straightforward:

(i) To explain the observed large anomaly in the specific heat of
URu$_2$Si$_2$, we need a rather strong order.

(ii) To be hidden, the strong order should have the symmetry of the
high-temperature phase, because otherwise it would be easily detectable by
x-ray and/or neutron diffraction.

(iii) If the order parameter has the symmetry of the high temperature
phase, the phase transition should be (contrary to experiments) of the
first order except the case of the time-reversal-symmetry breaking.
Therefore the most plausible candidate for the ``hidden order'' in URu$_2$Si$_2$
is a time-reversal-symmetry breaking system of magnetic moments with the
symmetry of the crystal lattice. In our version, this is the tetragonal
lattice of anti-toroidal vortices.

(iv) This conjecture has been fully confirmed in our {\em ab initio}
calculations.

Now we want to show that the ``hidden order'' of this type can be detected
means of careful monitoring of neutron reflections across the HO phase
transition. It is helpful that the lattice symmetry favors the ATV HO with
very unusual distributions of the intra-atomic magnetization resulting in
unusual form-factors for magnetic neutron scattering (see Fig.~\ref{Bragg}
for the reflection intensities obtained from ab-initio calculated
$\mathbf{M(r)}$). An obvious unusual feature is that high-symmetry
reflections $h00$, $0k0$, $00\ell$, and $hh0$ are zero for both the
ferro-vortex and antiferro-vortex phases. The main difference between two
phases is that there are pure magnetic reflections $h+k+\ell=2n+1$ in the
antiferro-vortex phase whereas for the ferro-vortex phase all the magnetic
reflections coincide with nuclear reflections $h+k+\ell=2n$. Comparison of
Fig. \ref{Bragg} with the observed intensities of pure magnetic
reflections \cite{Broholm1991} (100, 102, 201, 203, 106, and 300) allows
us to exclude the antiferro-vortex phase from the list of possible
candidates for HO.

The situation with the ferro-vortex phase is much more intriguing: the
magnetic reflections only slightly change the nuclear reflection
intensities; the latter have never been measured carefully for
URu$_2$Si$_2$ across the HO temperature. Moreover, the interference
between magnetic and nuclear contributions should vanish in the case of
equal fractions of clockwise and anticlockwise domains. According to our
calculations, the magnetic structure factor can reach its maximum
$\approx$0.25 $\mu_B$ for reflection 525 at $T=0$. However, this
reflection has a large nuclear structure factor. Fortunately, there are
many weak nuclear reflections with comparable magnetic factors from 0.15
to 0.2 $\mu_B$, for instance, 307 and 417; they are more sensitive to
magnetic scattering. It seems that accurate measurements of neutron
reflections as a function of temperature provide the only way to study ATV
HO quantitatively. Similar neutron experiments have revealed an unusual
magnetic order preserving translational symmetry of the lattice in the
enigmatic pseudogap phase of high-temperature superconductors
\cite{Fauque2006,Mook2008,Li2008,Li2011,Mangin2014}. We have found
recently a striking similarity between hidden orders in URu$_2$Si$_2$ and
in the pseudogap phase that will be discussed elsewhere. Quite probably,
the URu$_2$Si$_2$ HO phase is generic and similar phases where the order
remains undetected because of its high symmetry can exist in other
materials.

In conclusion, it is shown that high magnetic symmetry of URu$_2$Si$_2$
crystal can explain why its ``hidden order'' remains hidden for many
years. There is no spatial symmetry breaking in the HO phase transition
and solely the time-reversal symmetry is violated. Owing to their $4/mmm$
symmetry, uranium atoms have zero dipole and quadrupole moments, and the
first non-zero magnetic moment of the uranium vortex is the quadrupole
toroidal moment which can be used as an order parameter in the Landau
theory of the HO phase. The simulations suggest that the vortex magnetic
order of URu$_2$Si$_2$ is indeed energetically favorable and strong enough
to be detected by neutron diffraction.

\section*{Acknowledgements}

We are grateful to S.~A.~Pikin, M.~V.~Gorkunov, F.~de Bergevin,
G.~Beutier, R.~Caciuffo, S.~P.~Collins, M.~Kl\'eman, Y.~O.~Kvashnin,
N.~V.~Ter-Oganesyan, and I.~V.~Tokatly for useful discussions and
communications. This work was supported by the Ministry of Science and Higher Education within the State assignment FSRC ``Crystallography and Photonics'' RAS in part of symmetry analysis, and by the grant of Prezidium RAS No. I.11$\Pi$ in part of {\em ab initio} calculations.

\newpage
{\Large\bf Supplemental Material}

{\em Remarks on the Landau theory for time breaking order
parameters.---}Let us consider a possible form of the phenomenological
Landau-Ginzburg theory for the HO with $4/mmm$ symmetry. The general
discussion of the problem had been done in  \cite{Shah2000,Chandra2015}
and we will follow this works making only necessary changes. The free
energy includes two order parameters: a large HO parameter $\psi$ which
co-exists and interacts with the secondary antiferromagnetic order
parameter $m$.

The corresponding Landau theory is also briefly discussed with emphasis on
symmetry restrictions for possible terms in the free energy.

For $\psi$ order parameter we can use the values of $(M_v(000)\pm
M_v(\frac12,\frac12,\frac12))/2$ averaged additionally over a small
macroscopic volume; the signs $+/-$ correspond to
ferrovortex/antiferrovortex phases. In this case both order parameters
break time-reversal symmetry, and one can expect the type (A) theory
according to \cite{Shah2000} with a bilinear interaction term $g_A m\psi$.
However, for the ferro-vortex and antiferro-vortex phases this term is not
invariant under spatial symmetry transformations and therefore the
interaction term is $g_B m^2\psi^2$ (type (B) theory). In both types of
theory, the interaction with the external magnetic field $H$ should be
biquadratic, $H^2\psi^2$. As a result the external field $H$ cannot fix
the sign of $\psi$ even in the ferro-vortex phase. At present it is not
clear how to induce a single-domain state with this type of HO. We do not
discuss here the gradient terms which should be non-trivial because of the
tensor nature of ATV HO.

{\it Details of ab initio simulations.} Full relativistic
pseudopotentials, taking into account the spin-orbit interaction, should
be used for non-collinear magnetization. There is no such potentials for
uranium and ruthenium at the {\sc Quantum ESPRESSO} website \cite{QE} and
we used the norm conserving (nc) relativistic potentials from the web page
of the THEOS group at EPFL \cite{theos}. We tried several pseudopotentials
corresponding to different exchange-correlation functionals: Perdew--Wang
(pw91) \cite{Perdew1992,pw91}, the Perdew--Burke--Ernzerhof (pbe, revpbe
and pbesol) \cite{Perdew1996}, Wu--Cohen (wc) \cite{Wu2006}, Becke--Perdew
(bp) \cite{Becke1988,Perdew1986} types for generalized gradient
approximations (GGA) and the Perdew--Zunger (pz) \cite{Perdew1981}
pseudopotentials using the Local-Spin-Density Approximation (LSDA).
However, for reasonable computation time, the convergence of the
self-consistent iterations was reached only with the Perdew--Zunger full
relativistic pseudopotentials Z.rel-pz-n-nc.UPF where Z=U,Ru,Si.

Few minor technical details of simulations: Two U atoms at
$\pm(\frac14,\frac14,\frac14)$ positions; four Ru atoms at
$\pm(\frac14,\frac34,0)$ and $\pm(\frac34,\frac14,\frac12)$; four Si atoms
at $\pm(\frac14,\frac14,0.623)$ and $\pm(\frac14,\frac14,0.877)$ in the
unit cell with $a=b=4.112$ \AA\  and $c=9.538$ \AA. This setting was
selected for better visualization of results. A $12\times 12\times 6$
Monkhorst-Pack mesh \cite{Monkhorst} was used (it was proved that
$16\times 16\times 8$ gave practically the same results); the wavefunction
energy cutoff: 50 Ry; the charge-density energy cutoff: 200 Ry; the
Marzari--Vanderbilt smearing broadening \cite{Marzari} was fixed at 0.02
Ry. The starting wave functions are either random or atomic plus random.

It is important to start simulations from low-symmetry phases, for
instance orthorhombic, because this way a small orthorhombicity, sometimes
observed in the HO phase, would be automatically included into
consideration. Of course it would be better to start the self-consistent
procedure from completely non-symmetric magnetic structure so that the
final magnetic symmetry would appear as a result of minimization. However
such procedure is very resource demanding. Therefore initially we started
from the URu$_2$Si$_2$ structures with small distortions of the $Pmmn, n.
59$ space symmetry. We concluded that our simulations do not demonstrate
any residual orthorhombicity.

We also tried to start from a mixture of ATV HO and conventional
antiferromagnetic order. During the minimization procedure, the
antiferromagnetic order disappeared progressively and finally we had pure
ATV HO. This corresponds to destructive interaction of those order
parameters.

\end{document}